\documentclass[aps,prl,twocolumn,showpacs,letterpaper]{revtex4}
\usepackage{graphicx,dcolumn,longtable,epsfig}
\usepackage[usenames]{color}
\usepackage{amssymb}
\usepackage{amsmath}
\usepackage{epstopdf}
\usepackage{bm}

\input epsf.tex
\newcommand{\etal}{{\it et al.}}

\def\jpb#1#2#3{J.~Phys.~B:~{\bf #1},\ #2\ (#3)}

\def\jcp#1#2#3{J.~Chem.~Phys.~{\bf #1},\ #2\ (#3)}

\def\pra#1#2#3{Phys.~Rev.~A~{\bf #1},\ #2\ (#3)}

\def\prl#1#2#3{Phys.~Rev.~Lett.~{\bf #1},\ #2\ (#3)}
\def\sci#1#2#3{Science~{\bf #1},\ #2\ (#3)}

\def\etal{{\it et al.}}
\def\bea{\begin{eqnarray}}
\def\eea{\end{eqnarray}}
\def\be{\begin{equation}}
\def\ee{\end{equation}}

\begin{document}
\title{Ultracold giant polyatomic Rydberg molecules: coherent control of molecular orientation}

\author{Seth T.\ Rittenhouse and H. R.\ Sadeghpour}
\affiliation{ITAMP, Harvard-Smithsonian Center for Astrophysics, Cambridge, Massachusetts 02138}
\date{\today}

\begin{abstract}
We predict the existence of a class of ultracold giant molecules formed from trapped ultracold Rydberg atoms and polar molecules. The interaction which leads to the formation of such molecules is the anisotropic charge-dipole interaction ($a/R^2$).  We show that prominent candidate molecules such as KRb and deuterated hydroxyl (OD) should bind to Rydberg rubidium atoms, with energies $E_b\simeq 5-25$ GHz at distances $R\simeq 0.1-1 \ \mu$m. These molecules form in double wells, mimicking chiral molecules, with each well containing a particular dipole orientation. We prepare a set of correlated dressed electron-dipole eigenstates which are used in a resonant Raman scheme to coherently control the dipole orientation and to create cat-like entangled states of the polar molecule.
\end{abstract}


\maketitle

The ability to tailor and control interactions amongst species in a quantum gas is a primary motivation for topical interface research in atomic, molecular, optical, condensed matter and chemical physics. Ultracold polar molecules and Rydberg atoms share in this distinction for being ideally suited for such studies, as they provide a number of convenient control handles \cite{polar, KRb, rydberg, MagRydreview09,prl1}. Both polar molecules and Rydberg atoms can have large permanent and induced dipole moments, making them amenable to manipulation with external and optical fields via long-range van der Waals ($R^{-6}$) and anisotropic dipole ($R^{-3}$) forces. Design of synthetic many-body textures and simulations of realistic quantum phases are conditional on this level of control \cite{jak00,santos}.

It has been proposed that a Rydberg electron interacting via the zero-range Fermi pseudopotential with a nearby perturbing atom could bind into ultralong range Rydberg molecular states. Isotropic molecules with MHz binding energies, "butterfly" molecules with GHz binding energies and large (tens of Debye) permanent dipole moments, and  "trilobite" molecules with tens of MHz binding energies and ultralarge (kDebye) dipole moments \cite{trilobite,butterfly} were predicted to exist in ultracold traps. An experiment with ultracold Rydberg atoms in a dense rubidium magneto-optical trap (MOT) recently realized the existence of the istotropic utralong range Rydberg molecules \cite{pfau}. 

\begin{figure}[t]
\begin{center}
\includegraphics[width=3in]{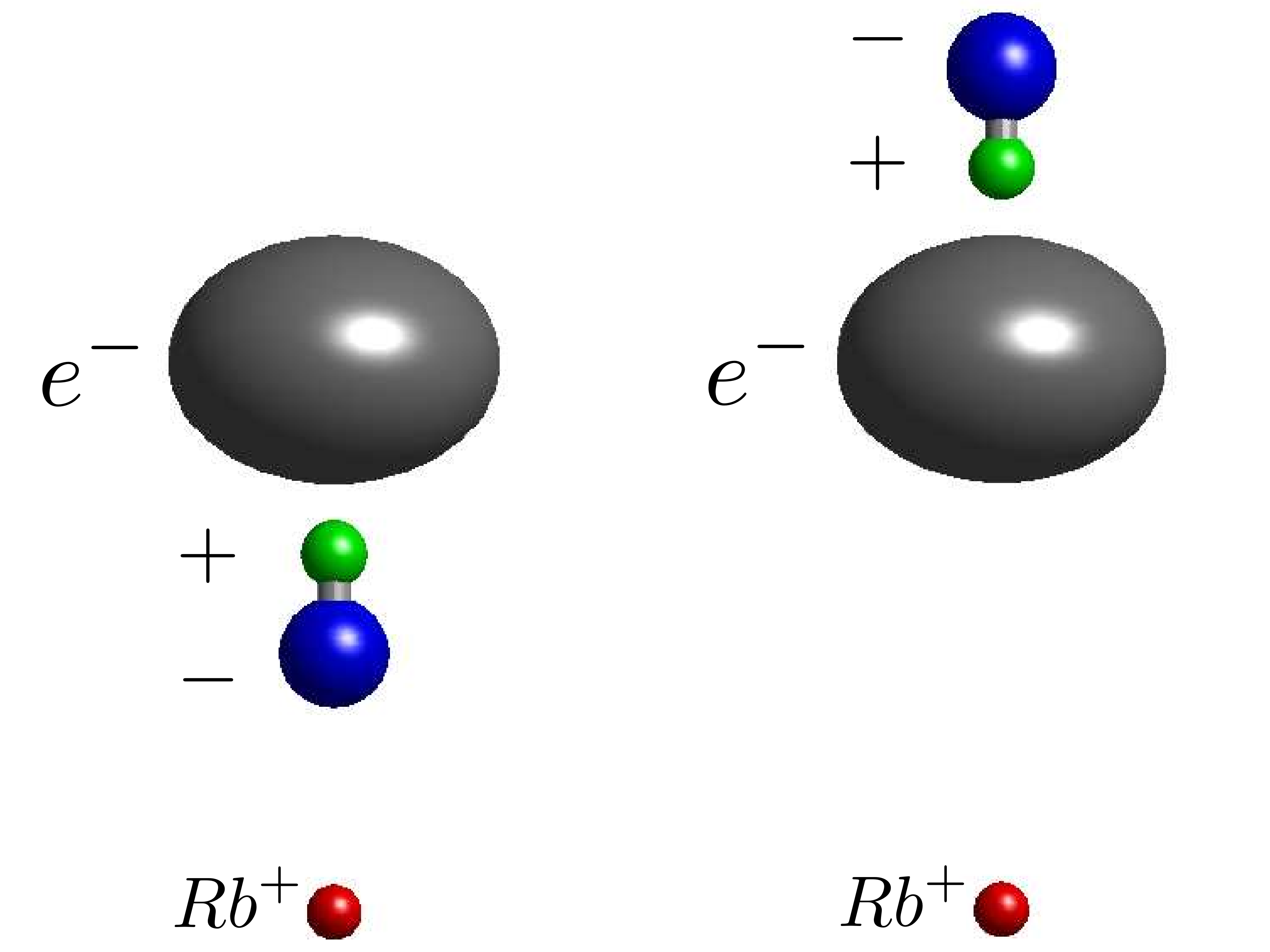}
\end{center}
\caption{A schematic of the two ultralong range Rydberg molecular states which form from the interaction of a Rydberg atom with a polar molecule. The Rydberg electron orbit is exaggerated to highlight the long-range nature of the interactions. The direction of the dipole is from the $+$ to $-$.}%
\label{scheme}%
\end{figure}

\begin{figure}[t]
\begin{center}
\includegraphics[width=3in]{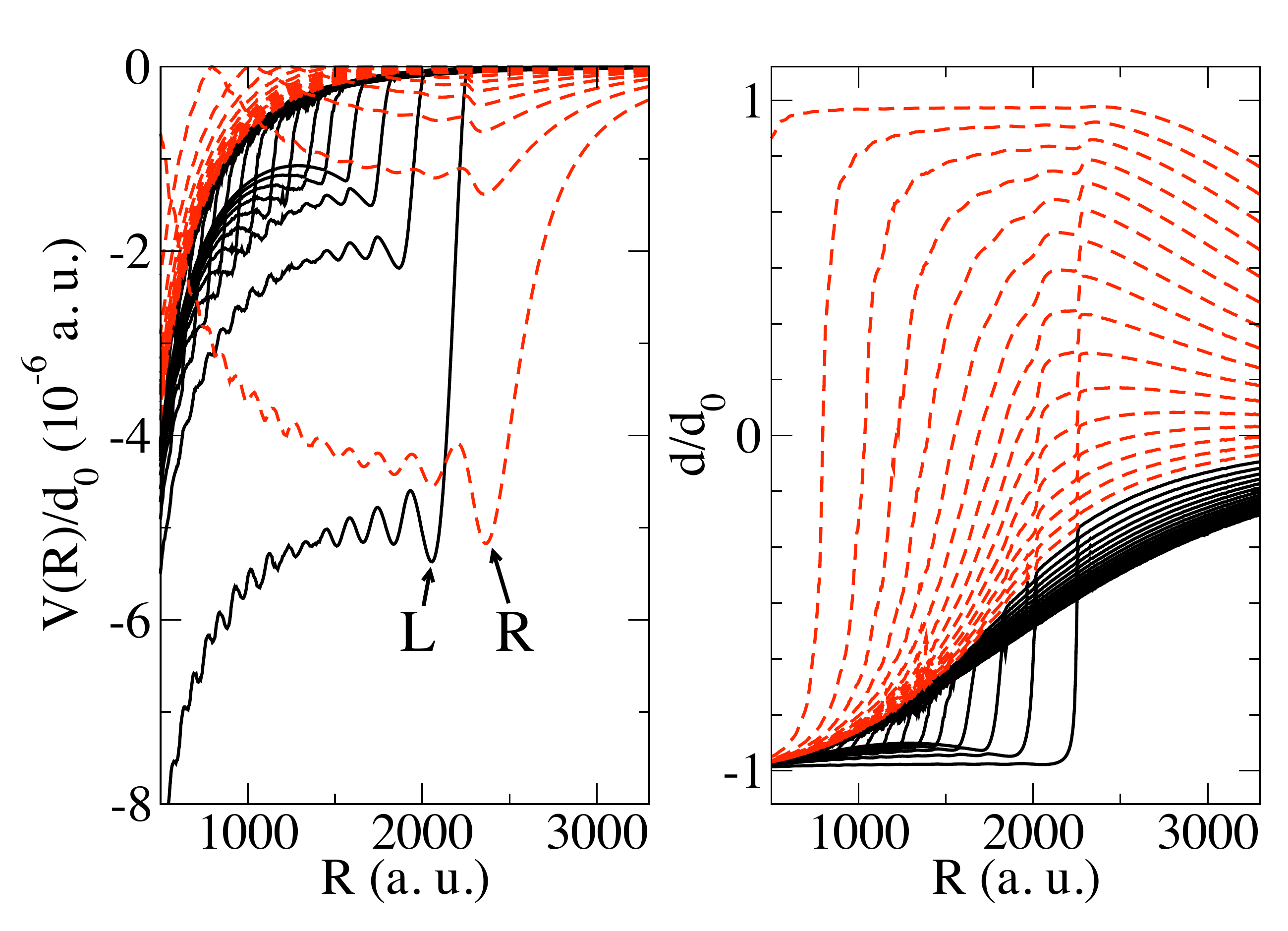}
\end{center}
\caption{The BO potential curves (left panel) for $^{87}$Rb(n=35)-OD($^2\Pi$) are shown. The zero of energy is the Rb(n=35) threshold. Only those angular momentum states with negligible quantum defect ($l \geq 3$) in Rb are included. The modulations in the potentials are due to the Rydberg oscillations and the avoided crossings are due to interactions in the degenerate manifold. The two lowest potential wells, near $R \sim 2000-2500$ a.u., are weakly coupled with the R (L) well supporting vibrational states of the Rb-OD molecule oriented away from (toward) the ion core. Positive values of the dipole (right panel) refer to the orientation of the molecular dipole away from the ion core and the negative values indicate the reverse.  }%
\label{n=35PotDip}%
\end{figure}

In this work, we propose the existence of a class of ultralong range macroscopic Rydberg molecules which can be created in ultracold traps. The molecular binding comes courtesy of the interaction of the Rydberg electron with the permanent dipole of the trapped polar molecule. This anisotropic interaction which has the form  $(a-\frac{1}{4}) /R^2$ is by far, the longest range interaction in the physics of cold neutral atoms and molecules and may open new vistas into the emergent discipline of long-range correlated quantum systems. These giant molecules have deep binding energies, $E_b \simeq 20$ GHz, are ultralong range, $R \simeq 2000 \ a_0$ (see Fig. \ref{n=35PotDip}), and are well protected from the non-adiabatic avoided crossings, which could lead to their premature demise. The double-well structure (left (L) and right (R) wells in Fig. \ref{n=35PotDip}), in which the orientation of the polar molecule dipole, $d$, is controlled by the Rydberg electron, resembles the double-well structure which determines the left- and right-handedness in a chiral molecule, as shown schematically in Fig. \ref{scheme}. We propose a realizable coherent Raman scheme to control the orientation of the molecular dipole and prepare cat-like superposition of the dipole states. 

The electron-dipole interaction $(a-\frac{1}{4}) /R^2$ becomes critical at $a=0$, resulting from a critical dipole, $d_{cr}=0.639315$ a.u. \cite{FT}. For $a>0$, the interaction is attractive ($d>d_{cr}$), and the electron binds to the dipole; for $a<0$, the dipole is subcritical ($d < d_{cr}$) and the electron does not bind to the dipole. When $d> d_{cr}=1.63D$, electron exchange occurs \cite{dipole}.

For the discussion outlined here, we treat the molecular dipole to be subcritical, $d < d_{cr}$, such that degenerate perturbation theory can be employed to obtain the Born-Oppenheimer (BO) potential energy curves. For repulsive interactions between the Rydberg electron and the molecular dipole ($d < d_{cr}$), details of the short range physics are unimportant. 

The starting point of our calculation is 
\be
V_{ed}\left(  \vec{R}-\vec{r}\right)  =\dfrac{\vec{d}\cdot\left(  \vec{R}%
-\vec{r}\right)  }{\left\vert \vec{R}-\vec{r}\right\vert ^{3}}%
\ee
where $\vec{R}$ is the dipole position with respect to the Rydberg core and
$\vec{r}$ is the electron position. If we assume that the dipole is on the $z$
axis, the matrix elements of this interaction in basis $|nlm>$ are%
\begin{widetext}
\be
\left\langle nlm\left\vert V_{ed}\right\vert nl^{\prime}m^{\prime
}\right\rangle   =\int d^{3}r\dfrac{d\left(  Z-r\cos\theta\right)
}{\left(  r^{2}+Z^{2}-2rZ\cos\theta\right)  ^{3/2}}
Y_{lm}^{\ast}\left(
\Omega\right)  Y_{l^{\prime}m^{\prime}}\left(  \Omega\right)  R_{nl}^{\ast
}\left(  r\right)  R_{nl^{\prime}}\left(  r\right)
\ee
With some algebra, the final form for the potential matrix becomes:
\begin{align}
\left\langle nlm\left\vert V_{ed}\right\vert nl^{\prime}m^{\prime
}\right\rangle  
=& d\delta_{mm^{\prime}}\sqrt{\left(  2l^{\prime}+1\right)  \left(
2l+1\right)  }
 \left(  -1\right)  ^{m}\sum_{l^{\prime\prime}=\left\vert
l-l^{\prime}\right\vert }^{l+l^{\prime}}\left(
\begin{array}
[c]{ccc}%
l & l^{\prime} & l^{\prime\prime}\\
0 & 0 & 0
\end{array}
\right)
\left(
\begin{array}
[c]{ccc}%
l & l^{\prime} & l^{\prime\prime}\\
m & -m & 0
\end{array}
\right) \\
 & \times \left[ \dfrac{ \left(  l^{\prime\prime}+1\right) }{Z^{l^{\prime
\prime}+2}}\int_{0}^{Z}r^{l^{\prime\prime}+2}R_{nl}^{\ast}\left(  r\right)
R_{nl^{\prime}}\left(  r\right)  dr 
- l^{\prime\prime}Z^{l^{\prime\prime}-1}%
\int_{Z}^{\infty}\dfrac{1}{r^{l^{\prime\prime}-1}}R_{nl}^{\ast}\left(
r\right)  R_{nl^{\prime}}\left(  r\right)  dr \right], \nonumber
\end{align}
\end{widetext}
where $R_{nl}(r)$ are the Laguerre polynomials and $\left(....\right)$ are $3-j$ symbols, and $Z={\vec R}\cdot {\hat Z}$.

A molecule becomes "polar", when the two opposite parity states of the molecule are mixed to create the permanent dipole moment $d_0$. The minimum electric field $F_c$ required to mix these states with a splitting $\Delta$ is $F_c=\frac{\Delta}{2d_0}$.  Defining the eigenvalues of the potential matrix as $V_e^\lambda (R)$, the BO  potentials for the interaction of a Rydberg atom with a polar molecule of dipole $d_0$ is

\be
V_{\lambda}\left(  R\right)  =d_0\left(  F_{c}-\sqrt{\left(  F_{e}^{\lambda
}\left(  R\right)  -\dfrac{1}{R^{2}}\right)  ^{2}+F_{c}^{2}}\right)
\ee
where $F_e^\lambda (R) = V_e^\lambda (R)/d$ is the electric field at position $R$ due to the Rydberg electron and $-\frac{d}{R^2}$ is the potential generated by the ion core. This post-diagonalization creates a set of dressed correlated electron-dipole eigenstates, with important implications for the control of the dipolar superposition.

The BO potential curves for Rb(n=35)OD($^2\Pi_{3/2}$) are shown in Fig. \ref{n=35PotDip}. Similar curves exist for Rb(n)KRb($^1\Sigma_{1/2}$) molecule. The modulations are due to Rydberg electron oscillations. The interaction of the ionic core with the dipole, $-\frac{1}{R^2}$, is seen in the potentials in which the molecular dipole points toward the core. The right panel in Fig. \ref{n=35PotDip} illustrates the transition of the molecular dipole orientation through the avoided crossings. The classification of the dipole moments according to their orientation is evident in this figure. The dashed curves show the evolution of the dipole which points away from the core to that which points toward the core (solid curves). The total molecular symmetry for the polyatomic molecule is either $^2\Sigma$ or $^3\Sigma$, depending on whether the polar molecule is $^1\Sigma$, as in KRb, or a doublet molecule, as in OD.

The formation of the giant Rydberg molecular state is a competitive
process between the electron- and core-dipole interactions. Binding
occurs when the electronic distribution is sufficiently concentrated,
allowing the electron-dipole interaction to overcome the core
contribution. The concentration of the electron wave function is in
part controlled by the range of $l$ values available in the degenerate
Rydberg manifold: $l=l_{\min},...,n-1$. If the minimum angular
momentum is too large compared to  $n$,
the electron interaction can not overcome the strong $1/R^2$ core
attraction and hence no binding occurs. We have found that for $l_{\min} >2$, $n\sim 20$ gives an
approximate cutoff below which no bound states can form.

\begin{figure}
\begin{center}
\includegraphics[width=3in]{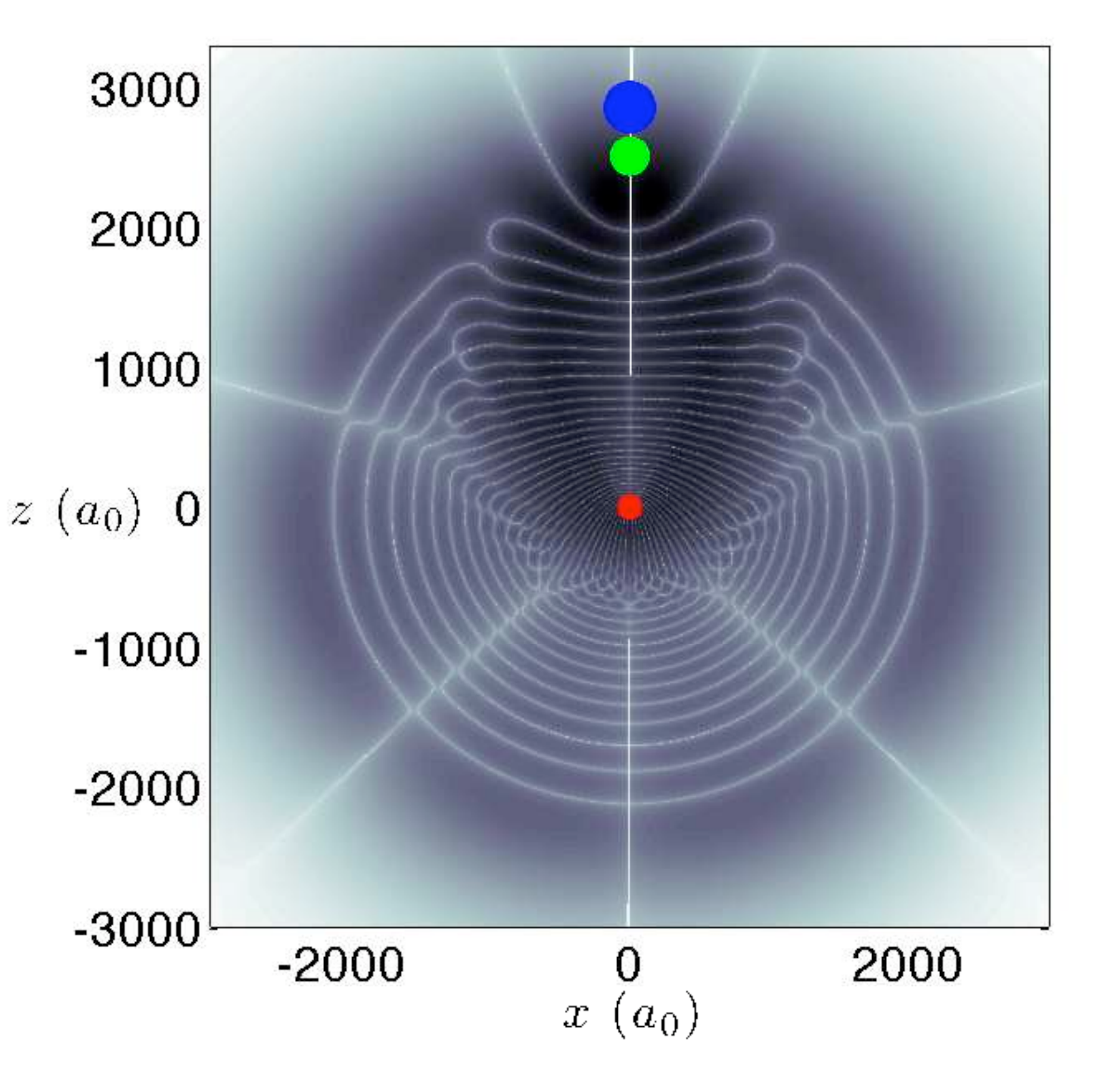}
\end{center}
\caption{(Color Online) A density plot of the electronic wave function for an
$n=35$ Rydberg molecule at $R=2300 a_0$
is shown in cylindrical coordinates. Darker areas show larger amplitudes of the electronic wave function.  The position of the $Rb^+$ (red circle) and the orientation of the molecular dipole (green-blue dots)
are indicated. The size of the diple is greatly exaggerated
for illustration. The spherical symmetry is broken by the anisotropic electron-dipole interaction.  This density plot is a realization of the right configuration in Fig. \ref{scheme}.
}%
\label{ryd_wf}%
\end{figure}

The electronic density in the xz-plane at $R=2300 \ a_0$ is shown in Fig. \ref{ryd_wf}. The orientation of the molecular dipole is at the top of the figure. The Rydberg oscillations are visible with large amplitude at enormous distances from the Rydberg core. The presence of a permanent molecular dipole distorts the spherical symmetry, illustrating the anisotropy of the electron-dipole interaction. The electron wave function seen in Fig. \ref{ryd_wf} clumps near the ``positive'' side of the dipole with a strong peak in the electron probability distribution.  

The size and energy scales of the giant molecules generally follow the Rydberg scaling with $n$. Specifically, the outer wells in  the lowest two R and L potentials have minima, respectively, at $R_e = 1.92(1) n^2$ and $R_e = 1.65(2) n^2$ with corresponding potential depths $V_R = -0.224(1)d_0/n^3$ and $V_L=-0.234(1)d_0/n^3$. For $d <d_{cr}$, the potential depth is less than 10\%  of  Rydberg energy spacing, allowing us to use degenerate perturbation theory. At $R_e$, the immense charge separation creates a large permanent dipole moment which scales as $d_{Rm} = 1.75n^2$- for $n=35$, it translates to  $d_{Rm}=5.5$kD.

\begin{figure}[t]
\begin{center}
\includegraphics[width=3.00in]{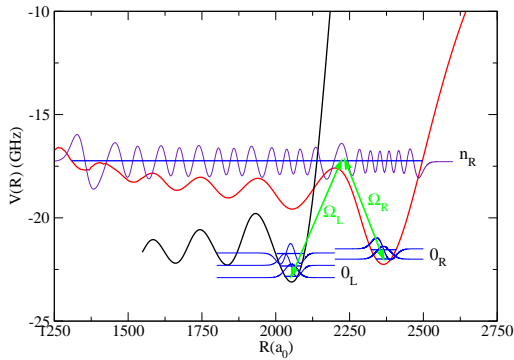}
\end{center}
\caption{The two (L and R) wells of the Rb-OD Rydberg molecule are shown. Also shown, are a select number of vibrational energy levels and associated wave functions, and our proposed Raman scheme for coherent population transfer between the two wells and preparation of cat-like states. The quantum numbers refer to the lowest vibrational states in the left ($0_L$), right ($0_R$), and an excited level in the right well ($v_R$), and $\Omega_L$, and $\Omega_R$ are the Rabi frequencies for the Raman transitions.}%
\label{TwoPots}%
\end{figure}

The two outer wells, each corresponding to the molecular dipole oriented toward or away from the ion core, are shown more clearly in Fig. \ref{TwoPots}. The two wells support a large number of vibrational levels within a typical transition frequency  in the $400-500$ MHz range. The coupling between the L and R curves is on the order of $O(\frac{\Delta^2}{ed_0})$, which for OD is $\simeq 10^{-15}$ a.u. The diagonal corrections to the energies in the $n=35$ manifold, are, respectively, $3\times 10^{-9}$ and $10^{-9}$ a.u.

The choice for a realizable polar molecule depends on two factors: such molecules should have subcritical dipoles, but be of sufficient strength for meaningful electron-dipole interaction, and have energy splittings which are a fraction of the Rydberg separation. The hydroxyl radical (OH), a $\Lambda$-doublet molecule ($^2\Pi_{3/2}$, $\Delta_\Lambda = 1.67$ GHz), was recently magnetically trapped \cite{OH} and its cold collision with Xe and He beams were studies \cite{AtomOH}.  Deuteration of hydrogen radicals reduces the doublet splitting by several factors \cite{BrownCarrington}, making them more readily polarizable. 

In Table \ref{t1}, we give a list of favorable polar molecules. The ground state of $^{40}$K$^{87}$Rb has been produced with large phase space density and its dipole moment and rotational splitting ($B_e$) have been established \cite{KRb}. While the dipole moment for the KRb molecule is small, its small rotational splitting ($2B_e$), and its ubiquity make it a good candidate. The deuterated radical, CD($ ^2\Pi_{3/2}$), can be polarized with field strength of $F_c\sim 2\times 10^{-7}$ a.u. and has a large but subcritical dipole moment, 1.46 D \cite{CH}. Another potentially realizable candidate is  metastable CO($a ^3\Pi_{1,2} (v=0)$) which lives for $2.63$ ms and its $\Delta_{\pm}$ transition ($\Omega=1, J=1$) is 394.1 GHz and has a dipole moment of 1.38 D \cite{CO}.

\begin{table}
\caption{\label{t1} Choice of polar molecules and their properties. \\  1 cm$^{-1}\simeq 30$ GHz.}
\begin{center}
\begin{tabular}
[c]{cccccc}\hline\hline
molecule & $d_{0}$(D) & $\Delta$(MHz) & $F_{c}$(a.u.) & $B_{e}$(cm$^{-1}$) &
Ref.\\\hline
OD $\left(  ^{2}\Pi_{3/2}\right)  $ & 1.66 & 494 & $6\times10^{-8}$ & $\sim40$
& \cite{OH} \\
KRb $\left(  ^{1}\Sigma_{g}\right)  $ & 0.566 & $-$ & $6\times10^{-7}$ &
$0.0371$ & \cite{KRb}\\
CD $\left(  ^{2}\Pi_{1/2}\right)  $ & 1.46 & 897 & $1.6\times10^{-7}$ &
$\sim60$ & \cite{CH}\\
CO $\left(  a^{3}\Pi_{1,2}\right)  $ & 1.38 & 394 & $7\times10^{-8}$ & $40$ & \cite{CO}
\\\hline\hline
\end{tabular}
\end{center}
\end{table}

The lowest vibrational levels in L ($0_L$) and R ($0_R$) wells in Fig. \ref{TwoPots} can be coherently coupled in a microwave Raman process. This fully on-resonance, but weak intensity two-photon transition, via an intermediate vibrational state ($v_R$) in the extended right well (dashed curve) allows for efficient transfer of population. The configurations in the L and R wells, respectively, are with the molecular dipole pointing toward or away from the ionic core. The coherent Raman scheme ensures that the molecular dipole orientations become entangled. Since the distance between the dipoles scales roughly as $0.27n^2$, which for $n=35$ translates to 16.5 nm, cat-like superposition states are possible. 

We compute the dipole transition strengths for the Raman process. Due to the existence of an inner wall in the potential curve which contains the R well, large Franck-Condon (FC) factors between the L and R states with the upper $v_R$ Raman states can be found. FC factors for $0_L\rightarrow v_R$ and $0_R\rightarrow v_R$ transitions are typically about 0.02 a.u. The electronic transition dipole moments, with photons polarized along the internuclear axis, scale as $0.20 n^2$;  for n=35 Rydberg state, the transition dipole has a value of  265.6 a.u., giving a total transition dipole, $d_{tr}\simeq 5$ a.u. To preserve the coherence of the cat states, $|c> = \frac{1}{\sqrt{2}}[|R> + e^{i \phi}|L>]$, in the presence of the polarization fields, on-resonance microwave field strengths of $F_{\mu} < F_c$ are desired. With $\Delta_\pm =2\Omega$, where  $\Omega={\bf d_{tr}}\cdot F_{\mu}$ is the Rabi frequency, microwave fields of $F_{\mu} \sim 10^{-7}$ a.u. (500 V/cm) can simultaneously maintain coherence in the cat states and drive the Raman transitions.

In summary, we explore the possibility that giant polyatomic Rydberg molecules could readily form in ultracold trapped mixture of atoms and molecules, from interaction of Rydberg atoms and polar molecules. These molecules which form in double wells contain dipole configurations which can be coherently controlled by a Raman process. The dipolar interaction of two polarized giant Rydberg molecule could be coupled to the internal cat states of the polar molecules. Ionization of the Rydberg atom from each L and R well can deterministically measure the orientational state of the polar molecule.

We are grateful to M. Cavagnero, C. Ticknor and J. Feist for discussions. This work was funded by a grant from the NSF to ITAMP at the Harvard-Smithsonian Center for Astrophysics and Harvard Physics Department.

\end{document}